\documentclass[pra,aps,twocolumn,superscriptaddress,amsmath,amssymb,nofootinbib]{revtex4}
\usepackage{graphicx}% Include figure files
\usepackage{epstopdf}
\usepackage{units}

\newcommand{\bra}[1]{\left<#1\right|}
\newcommand{\ket}[1]{\left|#1\right>}

\begin{document}

\title{Arbitrary quantum-state preparation of a harmonic oscillator
  via optimal control}
\date{\today}

\author{Katharina Rojan}
\affiliation{Theoretische Physik, Universit\"at des Saarlandes, 
  D-66123 Saarbr\"ucken, Germany}
\author{Daniel M. Reich}
\affiliation{Theoretische Physik, Universit\"at Kassel,
  Heinrich-Plett-Str.40, D-34132 Kassel, Germany}
\author{Igor Dotsenko}
\affiliation{Laboratoire Kastler-Brossel, ENS, UPMC-Paris 6, CNRS, coll\`ege de France, 24
  rue Lhomond, 75005 Paris, France} 
\author{Jean-Michel Raimond}
\affiliation{Laboratoire Kastler-Brossel, ENS, UPMC-Paris 6, CNRS, coll\`ege de France, 24
  rue Lhomond, 75005 Paris, France} 
\author{Christiane P. Koch}
\affiliation{Theoretische Physik, Universit\"at Kassel,
  Heinrich-Plett-Str.40, D-34132 Kassel, Germany} 
\author{Giovanna Morigi}
\affiliation{Theoretische Physik, Universit\"at des Saarlandes,
  D-66123 Saarbr\"ucken, Germany}

\begin{abstract}
  The efficient initialization of a quantum system is a prerequisite for
  quantum technological applications. Here we show that several
  classes of quantum states of a harmonic oscillator can be efficiently
  prepared by means of a Jaynes-Cummings interaction with a single
  two-level system. This is achieved by suitably tailoring external
  fields 
  which drive the dipole and/or the oscillator. The time-dependent
  dynamics that leads to the target state is identified by
  means of Optimal Control Theory (OCT) based on Krotov's method. Infidelities
  below $10^{-4}$ can be reached for the parameters of the
  experiment of the ENS group in Paris, where the oscillator is a mode of a high-Q
  microwave cavity and the dipole is a Rydberg transition of an atom.
 For this specific situation we analyze the limitations on the fidelity due to 
 parameter fluctuations and identify robust dynamics based on pulses found using ensemble OCT. 
 Our analysis can be extended to quantum-state preparation of continuous-variable systems in other platforms, such as trapped ions and circuit QED. 
\end{abstract}

\maketitle

\section{Introduction}
Control of the quantum dynamics of physical systems lies at the core
of quantum technological applications. A key issue is the 
initialization of nonclassical states 
which requires a sufficiently high fidelity to
permit efficient information processing. In this context,
protocols based on quantum optimal control theory~\cite{SomloiCP93,ZhuJCP98}
have been acquiring increasing relevance. This is due to 
the flexibility of the approach, that
allows for implementing almost arbitrary
dynamics~\cite{PalaoPRL02,TeschPRL02,PalaoPRA03} over relatively short time scales, 
see Refs.~\cite{DeChiaraPRA08,TommasoPRL09,MurphyPRA10,GoerzJPB11} for
a few examples.

Here we apply Optimal Control Theory (OCT) based on Krotov's
method~\cite{Konnov99,SklarzPRA02,Reich} 
to the efficient preparation of the quantum state of a
harmonic oscillator, which interacts for a fixed time with a dipolar
transition. The dipolar transition is quasi-resonant with the
oscillator frequency and couples to it via a Jaynes-Cummings type of
dynamics \cite{JaynesCummings}. This coupling renders the harmonic oscillator controllable. With the proper sequence of pulses, it is possible to perform any desired unitary transformation on the Hilbert space spanned by the dipole states together with the lowest $n$ energy levels of the oscillator~\cite{LawEberly,BlochRangan,HuanLloyd}. 
Specific implementations of 
algorithms based on OCT of the Jaynes-Cummings
dynamics include quantum state preparation of a trapped ion's center
of mass motion \cite{Ben-Kish,RanganMonroe}  and of superconducting
circuits  \cite{Hofheinz2008,Hofheinz2009}. Here, we focus on quantum
state preparation of the electromagnetic-field mode of a high-finesse
microwave resonator via the interaction with a transition of a Rydberg atom. 
Our purpose is to theoretically analyze the efficiency of
quantum state preparation of a class of nonclassical states, which
have been often discussed in the literature. The efficiency of most of the proposed protocols for these states is limited by the onset of decoherence and by the fact that, in some cases, they are based on projective measurements. 
It is thus desirable to identify generic procedures for identifying {\it deterministic} protocols
which can be realized over sufficiently {\it fast times} to avoid the
detrimental effect of decoherence. We address these issues by developing optimal-control based protocols.

\begin{figure}[bt] 
  \begin{center}  
    \includegraphics[width=1\linewidth]{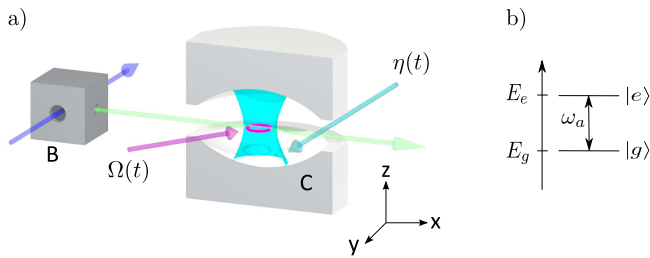} 
  \end{center}
   \caption{\label{Fig:1} Scheme of the considered experimental setup. (a) The field of a high-finesse microwave cavity \textsf{C} is prepared in an arbitrary quantum state by means of the interaction with a quasi-resonant dipolar transition of a circular Rydberg atom, schematically shown in (b). The arbitrary superposition of atomic states $\ket{e}$ and $\ket{g}$ is prepared in \textsf{B}. The atom then crosses  \textsf{C} with a fixed velocity, defining the interaction time. The target field state is reached with high fidelity by a suitably tailored time-dependent dynamics identified by means of OCT. The latter delivers the time-dependent form of the electric fields driving the atom, at coupling strength  $\Omega(t)$, and the cavity, at strength $\eta(t)$, that efficiently drive the cavity mode into the target state.}
\end{figure}

Our theoretical analysis makes specific reference to the setup of the experiment in the ENS group in Paris
\cite{Haroche_RMP,HarocheBook}. The elements of the experiment, which
are relevant to 
our study, are schematically illustrated in Fig.~\ref{Fig:1}: A
mode of the electromagnetic field is driven by the dipolar transition
between two circular-Rydberg states of an atom flying through the
resonator. Atom and microwave field mode undergo a textbook realization
of the Jaynes-Cummings 
dynamics \cite{JaynesCummings,HarocheBook}. A fixed interaction time
is set by selecting the atom's velocity. To this setup we add
the control tools, which are classical fields driving the atomic
transition and/or the cavity mode. The specific shape of the fields is
determined by OCT using Krotov's method. Our
target is the realization of specific quantum states of the
resonator with fidelities exceeding $0.9999$ (and correspondingly infidelities below $10^{-4}$).  

The time-dependent Hamiltonian we optimize has first been analyzed by
Law and Eberly~\cite{LawEberly} specifically for arbitrary
control of the mode of a quantum electromagnetic field. 
They proposed a procedure based on a sequence of unitary
transformations, in which only one coupling element of the Hamiltonian
acts at a time. This protocol has the asset to offer 
physical insight into the dynamics one needs to shape and serves as
initial guess to our search.  Our procedure, and optimal control in
general, goes nevertheless beyond this intuitive procedure, as it simultaneously considers all
processes driving the system and thus exploits interference among
them.  Analysis performed with superconducting circuits demonstrate
its better performance over Law-Eberly type of schemes~\cite{Moelmer}.  

The article is organized as follows. Section~\ref{Sec:Model}
introduces the Hamiltonian and the optimization
algorithm. Specifically, we discuss the functional which is minimized 
by means of OCT and the different classes of target
states. The results for quantum state preparation  are presented in 
Sec.~\ref{Sec:results}. We show how experimental uncertainties and
noise can be accounted for in Sec.~\ref{sec:exp}
and draw our conclusions in Sec.~\ref{Sec:conclude}.

\section{Model}
\label{Sec:Model}

The physical system we consider is a harmonic oscillator
that is coupled to a quasi-resonant two-level transition
(dipole) by a Jaynes-Cummings type of interaction. Our objective is to
prepare the harmonic oscillator in a specific desired state at the end
of a fixed interaction time $\tau$ with the dipole.  This is achieved by means of optimized electric fields, which either 
couple to the atom via a side propagation through the cavity, and/or to the cavity via a diffraction on the mirrors edges or surface defects. 

\subsection{Time-dependent Hamiltonian}

Let $\omega_f $ be the frequency of the oscillator and $a,a^\dagger$
the annihilation and creation operators of an oscillator excitation, with
$[a,a^\dagger]=1$. We denote by $|n\rangle$ the number-state basis of
the harmonic oscillator, with $n=0,1,2,\ldots$, such that
$a|n\rangle=\sqrt{n}|n-1\rangle$ for $n>1$ and $a|0\rangle=0$. The
oscillator couples to a dipolar transition at frequency $\omega_a$
with  ground state $|g\rangle$ and the excited state $|e\rangle$, as illustrated in Fig. \ref{Fig:1}(b).  
The dynamics is generated by the time-dependent Hamiltonian $H(t)$,
\begin{equation}
  \label{Hges}
  H(t) = H_0 + H_{c}(t)\,.
\end{equation}
Here, $H_0$ is the Jaynes-Cummings Hamiltonian, which governs the
coupled oscillator-atom dynamics in absence of external drives. It
reads 
\begin{equation}\label{Hdrift}
  H_0 = \hbar \omega_f a^{\dagger}a + \hbar\omega_a\frac{\sigma_z}{2}
  +\hbar \frac{g}{2}(a^{\dagger}\sigma+\sigma^{\dagger}a)\,,
\end{equation}
with $\sigma_z$ the Pauli operator for the dipole pseudo-spin and $g$ the vacuum-Rabi frequency, determining the strength of the
coupling between oscillator and dipole. For $\omega_a=\omega_f$ the
eigenstates of the Jaynes-Cummings Hamiltonian \eqref{Hdrift} are the
dressed states $\ket{\pm,n}$, which read
\begin{eqnarray}
  \ket{\pm,n}=(\ket{g,n+1}\pm\ket{e,n})/\sqrt{2}\,,
\end{eqnarray}
with eigenvalues $E_{\pm,n}=\hbar( n \omega_f\pm g\sqrt{n})$. The
coupling between dipole and oscillator gives thus rise to an anharmonic
spectrum of excitations of the composite system.  

The Hamiltonian $H_{c}(t)$ contains the coupling to the external
fields which can induce a time-dependent Stark shift $\Delta(t)$
on the dipolar transition. It can also quasi-resonantly drive the atomic
transition and cavity mode with time-dependent strengths
$\Omega(t)$ and $\eta(t)$, respectively: 
\begin{eqnarray}
  \label{Hctrl}
  H_{c} &=& \hbar\Delta(t)\frac{\sigma_z}{2}+
  \hbar\frac{\Omega(t)}{2}(e^{-i(\phi_l(t)+\omega_lt)}{\sigma}^{\dagger}+{\rm H.c.})\nonumber\\
  &&+\hbar\frac{\eta(t)}{2}(e^{-i(\phi_p(t)+\omega_p t)}a^{\dagger}+{\rm H.c.})\,.  
\end{eqnarray}
The 
pulses driving cavity and atom have carrier frequencies  $\omega_p$ and
$\omega_l$, respectively, with corresponding phases $\phi_p$ and $\phi_l$.
To reduce the numerical effort, 
we employ a frame that rotates with the oscillator frequency. The
total Hamiltonian then reads
\begin{subequations}
\begin{eqnarray}\label{Hrichtig}
 H^{'}(t)&=& \hbar(\omega_a-\omega_f+\Delta(t))\frac{\sigma_z}{2}+ \hbar \frac{g}{2}(a^{\dagger}\sigma +\sigma^{\dagger}a)\nonumber \\ &&+\hbar\left(\frac{\tilde{\Omega}(t)}{2}\sigma^{\dagger}+\frac{\tilde{\eta}(t)}{2}a^{\dagger}+{\rm H.c.}\right)\,,
\end{eqnarray}
with complex-valued controls 
\begin{eqnarray}
&&\tilde\Omega(t)=\Omega(t)e^{-i(\omega_l-\omega_f) t}e^{-i\phi_l(t)}\,,\\
&&\tilde\eta(t)=\eta(t)e^{-i(\omega_p-\omega_f)  t}e^{-i\phi_p(t)}\,.
\end{eqnarray} 
\end{subequations}
Equation~\eqref{Hrichtig} accounts for an imposed time-dependent Stark shift of the atomic transition, which are generated by a pulse of amplitude $\vartheta(t)$. We report it for completeness, since the realizations we consider here will not implement this latter kind of pulses \footnote{These pulses can take the form $\Delta(t)=\vartheta^2(t)/\delta$ where $\vartheta(t)$ is the
slowly-varying amplitude of a third external field coupling state
$\ket{g}$ to an auxiliary state $\ket{h}$ at detuning
$|\delta|\gg|\max_t(\vartheta(t))|$. In the ENS experiment this can be more simply realized by using the differential Stark effect in a static electric field.}. 

We seek to identify optimized time dependences of the classical
fields which efficiently lead to the preparation of
a target state of the oscillator, denoted by $\ket{\Phi_{\rm target}}$, starting from a well
defined initial state $\ket{\phi(0)}$ of oscillator and dipole. In
particular, we will assume that the oscillator is initially in the
ground state, $\ket{\Phi(0)}=\ket{0}$,
whereas we take the atom to be in the most convenient
state $\ket{\Psi_a(0)}$, depending on the target. 
The initial state thus reads 
\begin{equation}
\label{phi:0}
\ket{\phi(0)}= \ket{\Psi_a(0)}\otimes\ket{0}\,,
\end{equation}
where
$\ket{\Psi_a(t=0)} = \alpha \ket{g} + \beta \ket{e}$, with $\alpha =
\cos\theta, \beta = e^{i\phi}\sin\theta$ and $\phi,\theta \in
\mathbb{R}$. The desired state at time $t=\tau$ has the form 
\begin{equation}
\label{phi:tau}
\ket{\phi(\tau)}= \ket{\Psi_a(\tau)}\otimes\ket{\Phi_{\rm target}}\,,
\end{equation}
where we do not impose any constraint on the atomic state
$\ket{\Psi_a(\tau)}$, except for the fact that atom and
cavity must be disentagled at time $t=\tau$.

\subsection{Optimal control theory}
 
In order to determine the classical controls which transfer
state~\eqref{phi:0} into 
state~\eqref{phi:tau} by the unitary time evolution that is generated
by Hamiltonian~\eqref{Hrichtig}, %{Hges},  
we minimize the functional  $J$,  which is composed of two terms: %which is defined as
\begin{equation}
  \label{functional}
  J=J_{\tau}+\int_{0}^{\tau} 
  J_t[\vartheta(t),\tilde{\Omega}(t),\tilde{\eta}(t)] \ dt\,.
\end{equation}
The first term on the right-hand side is the final time
functional which corresponds to the infidelity, i.e., the 
difference between unity and the fidelity for obtaining the target
state $\ket{\Phi_{\rm target}}$, 
\begin{equation}
  J_{\tau} 
  = 1-\bra{\Phi_{\rm target}}
  {\rm Tr}_{\mathrm{a}}\left[U(\tau)\rho(0)U^{\dagger}(\tau)\right]\ket{\Phi_{\mathrm{target}}}\,.
\end{equation}
Here, $U(\tau)=\mathcal{T}\{\exp(-{\rm i}\int_0^\tau {\rm d}tH(t))\}$ is the
unitary evolution operator generated by $H(t)$ in Eq.~\eqref{Hges}, with
$\mathcal{T}$ being the time-ordering operator, while
$\rho(0)$ denotes the initial state of the total system, 
$\rho(0)=\ket{\phi(0)}\bra{\phi(0)}$, and ${\rm Tr}_{\rm a}$ is
the partial trace over the dipolar degrees of freedom. The
presence of the partial trace indicates that the 
fidelity for reaching the cavity target state is optimized regardless
of the dipole final state. This ideally requires the dipole to be
disentagled from the oscillator at time $t=\tau$.  

The second term in Eq. \eqref{functional} is the intermediate-time
functional.  
It explicitly depends on the controls $\vartheta(t)$ (i.e., the field
that generates the detuning $\Delta(t)$), 
$\tilde{\Omega}(t)$, and $\tilde{\eta}(t)$. A convenient choice
corresponds to minimizing the change of the controls with respect to
reference fields~\cite{PalaoPRA03}, 
\begin{eqnarray}
  J_{t}[\vartheta(t),\tilde{\Omega}(t),\tilde\eta(t)]
  &=&\frac{\lambda_{\vartheta}}{S_{\vartheta}(t)}[\vartheta(t)-\vartheta_{ref}(t)]^2\nonumber\\ 
  &&+\frac{\lambda_{\tilde\Omega}}{S_{\tilde\Omega}(t)}[\tilde{\Omega}(t)-\tilde\Omega_{ref}(t)]^2
  \nonumber \\
  && +\frac{\lambda_{\tilde\eta}}{S_{\tilde\eta}(t)}[\tilde{\eta}(t)-\tilde\eta_{ref}(t)]^2\,.
\end{eqnarray}
Here, $S_{\vartheta}(t)$, $S_{\tilde\Omega}(t)$ and $S_{\tilde\eta}(t)$ are shape
functions to ensure a smooth switch on and off of the control fields at times
$t=0$ and $t=\tau$. Unless specified otherwise, we take them to be
$\sin^2(\pi t/\tau)$, %which is equal to one for most times $t$,
allowing maximum flexibility for shaping $\vartheta(t)$,
$\tilde\Omega(t)$ and $\tilde\eta(t)$. The parameters
$\lambda_{\vartheta}$, $\lambda_{\tilde\Omega}$ and $\lambda_{\tilde\eta}$
represent weights that govern the step size of the optimization, and 
$\vartheta_{ref}(t)$, $\tilde\Omega_{ref}(t)$, and
$\tilde\eta_{ref}(t)$ are reference fields. A good choice takes the
references fields to be the controls obtained from the previous step
of the iterative optimization. This ensures that $J_t$ tends to zero
as the optimum is approached such that the value of the total
functional $J$ close to the optimum is solely determined by the
final-time part $J_\tau$~\cite{PalaoPRA03}. 

An optimization problem is completely specified in terms of the
optimization functional, equations of motion, and coupling to the
controls~\cite{Reich}. Based on these ingredients, Krotov's method
allows for deriving an optimization algorithm that, in the continuous
time limit, guarantees monotonic convergence~\cite{Konnov99}. It
consists in the coupled control equations which need to be solved
iteratively. An implementation of Krotov's method is found for example
in the spin dynamics software spinach~\cite{HogbenJMR11,spinach}. 
In our example the linear version of Krotov's method is
sufficient for obtaining a monotonically convergent
algorithm, since we deal with linear equations of motion, with an intermediate-time
functional that is independent of the state of the system, and with a
final-time functional that depends only linearly on the state of the
system~\cite{Reich}. The update equation for the
control, exemplarily given for $\tilde\Omega(t)$, reads
\begin{widetext}
  \begin{eqnarray}
    \tilde\Omega^{(i+1)}(t)=\tilde\Omega^{(i)}(t)
    +\frac{S_{\tilde\Omega}(t)}{\lambda_{\tilde\Omega}} \, 
    \mathfrak{Im}\bigg\{\bra{\chi^{(i)}(t)}
    \frac{\partial H}{\partial \tilde\Omega}\bigg|_{\tilde\Omega^{(i+1)}(t),\varphi^{(i+1)}(t)}
    \ket{\phi^{(i+1)}(t)}\bigg\}  \,, \label{eq:update}
  \end{eqnarray}  
\end{widetext}
where $i$ denotes the iterative step.
The equations for $\tilde\eta(t)$ and $\vartheta(t)$ are completely
analogous to Eq.~\eqref{eq:update}. 
Calculating the improved control $ \tilde\Omega^{(i+1)}(t)$ according
to Eq.~\eqref{eq:update} requires
forward propagation of the state of the system $\ket{\phi}$ under
the 'new' controls, $\tilde\Omega^{(i+1)}(t)$, $\tilde\eta^{(i+1)}(t)$,
$\vartheta^{(i+1)}(t)$, 
and backward propagation of the so-called adjoint states $\ket{\chi}$
under the 'old' controls  $\tilde\Omega^{(i)}(t)$, $\tilde\eta^{(i)}(t)$,
$\vartheta^{(i)}(t)$. That is, $\ket{\phi^{(i+1)}(t)}$ and
$\ket{\chi^{(i)}(t)}$ are obtained as 
the solution of the Schr\"odinger equations
\begin{equation}
    \frac{d}{dt}\ket{\phi^{(i)}(t)} =-\frac{\mathrm{i}}{\hbar} 
    H[\vartheta^{(i+1)}(t),\tilde\Omega^{(i+1)}(t),\tilde\eta^{(i+1)}(t)]
    \ket{\phi^{(i)}(t)}\,,
\end{equation}
and 
\begin{subequations}
  \label{eq:backward}
  \begin{equation}\label{eq:backwardSE}
    \frac{d}{dt}\big|\chi^{(i)}(t)\big\rangle=-\frac{\mathrm{i}}{\hbar}
    H[\vartheta^{(i)}(t),\tilde\Omega^{(i)}(t),\tilde\eta^{(i)}(t)]
    \big|\chi^{(i)}(t)\big\rangle\,. 
  \end{equation}
For our specific choice of optimization functional and Hamiltonian, 
the equation of motion for the adjoint state
$\ket{\chi}$, Eq.~\eqref{eq:backwardSE}, 
turns simply out to be the standard Schr\"odinger
equation~\cite{Reich}.  
The initial condition for the backward propagation, 
at final time $\tau$, is given by the derivative of 
the final-time functional with respect to $\bra{\phi}$, evaluated at
time $t=\tau$:
\begin{equation}
  \big|\chi^{(i)}(t=\tau)\big\rangle=-\nabla_{\bra{\phi}} 
  \left.J_{\tau}\right|_{\ket{\phi^{(i)}(t=\tau)}}\,.
\end{equation}
For $J_\tau$ in Eq.~\eqref{functional}, it becomes 
\begin{equation}
  \big|\chi^{(i)}(\tau)\big\rangle=   
  \sum_a \ket{a,\Phi_{\mathrm{target}}}\bra{a,\Phi_{\mathrm{target}}}\phi(\tau)\rangle \,,
\end{equation}
\end{subequations}
where the states $\ket{a}$ correspond to an orthonormal basis of the dipole's Hilbert space.
Since Eq.~\eqref{eq:update} is implicit in
$\tilde\Omega^{(i+1)}(t)$, the time grids for the states and for the
controls are shifted by $\Delta t/2$ such that
$\tilde\Omega^{(i+1)}(t+\Delta t/2)$ is obtained from
$\ket{\phi^{(i+1)}(t)}$~\cite{PalaoPRA03}. 
The iteration is started by choosing a guess for each of the
controls. The guess fields must be physically sensible
choices, otherwise the change in the control in Eq.~\eqref{eq:update}
may be very small and convergence correspondingly slow. 

\subsection{Targets}
\label{subsec:targets} 

Our goal is to determine controls for the 
preparation of the following classes of states of the harmonic oscillator, namely,
(i) Fock states of arbitrary number $n$,  
\begin{equation}
  \ket{\Phi_{\rm target}}=\ket{n}\,,\label{Fock}
\end{equation}
(ii) Fock state superpositions of the form
\begin{equation}
  \ket{\Phi_{\rm target}}=\frac{\ket{0}+\ket{n}}{\sqrt{2}}, \quad
  \text{with} 
  \quad n>1\,, \label{Focksup}
\end{equation}
and (iii) even cat states such as
\begin{equation}
  \ket{\Phi_{\rm target}}=\ket{\Phi^+_{\rm cat}}
  =\frac{\ket{\alpha}+\ket{-\alpha}}{\sqrt{2(1+e^{-2|\alpha|^2})}} \,,\label{cat}
\end{equation}
where
$|\alpha\rangle=e^{-|\alpha|^2/2}\sum_n\alpha^n/(\sqrt{n!})|n\rangle$
is a coherent state and $\alpha$ a complex number.  
The restriction to these classes of states is motivated by the
possibility of comparing the efficiency and the dynamics under the
optimized controls to  previous work which employed protocols based on
ingenious intuition of the key elements leading to the desired target
states~\cite{Davidovich1996,Haroche_RMP,Monroe1996,Leibfried,Ben-Kish,Hofheinz2009,RanganMonroe,KneerLaw,Domokos}.  
In this way we can acquire a better understanding of the dynamics induced by the optimal control protocol we identify. 

There are several physical platforms in which the dynamics 
discussed here can be implemented. We focus on microwave cavity
quantum electrodynamics (QED), where the harmonic oscillator is a mode of a
high-finesse microwave resonator and the dipole is a quasi-resonant
atomic transition between two Rydberg states.  
The protocols that have so far been implemented experimentally  in
microwave cavity QED are based on the interaction of the cavity field
with a beam of atoms~\cite{Walther,Haroche_RMP} and in several cases
rely on projective measurements. Our purpose is to develop protocols
which lead to the {\it deterministic} preparation of an arbitrary
state of the microwave field with a {\it single} atom.   

Before presenting our results we recall existing proposals for and
implementations of the preparation of Fock states, Fock state
superpositions, and Schr\"odinger cat states.
Several proposals for preparing the mode of a resonator in a Fock state can be found in the literature, see e.g.
Refs.~\cite{Filipowicz,Slosser,Walther,Davidovich1996,Domokos,Haroche_RMP,Solano2001,KimblePhotonBlockade,RempePhotonBlockade}. Fock states have been experimentally realized with the motion
of single trapped ions coupled via lasers to an ion internal
transition~\cite{Leibfried}, with a high-Q mechanical resonator
coupled to superconducting circuits~\cite{Hofheinz2008,Hofheinz2009},
in circuit QED~\cite{Pechal2014}, and with the mode of a high-Q cavity
coupling to a two-level
transition~\cite{Rempe,Brune,Bertet,Nogues,Weidinger,Varcoe,Sayrin,Walther,KimbleJPB,Keller2004,Kimble1photon,SinglePhotonMPQ,KimblePhotonBlockade,RempePhotonBlockade}.   
Fock state superpositions, cf. Eq.~\eqref{Focksup}, have been
deterministically created with superconducting
circuits~\cite{Hofheinz2008,Hofheinz2009}. The ability of creating
such states is required in order to prepare so-called NOON states of
two harmonic oscillators~\cite{Wilhelm}, which are 
relevant for quantum metrology. 
Schr\"odinger cat states are popular in the literature due to their peculiar
nonclassical properties. Remarkable experimental realizations have been reported in microwave cavity
QED~\cite{Haroche_Cat,Haroche_Cat2}, trapped
ions~\cite{Monroe1996,Leibfried,Roos}, and most recently with 
superconducting circuits~\cite{Hofheinz2009}. 

The efficiency of most of these protocols is limited by the onset of
decoherence and by the fact that, in some cases, they are based on projective measurements. 
It is thus desirable to identify generic procedures for identifying deterministic protocols
which can be realized over sufficiently fast times to avoid the
detrimental effect of decoherence.   

\section{Results}	
\label{Sec:results}

The time scale $\tau$ of the interaction is fixed in relation to the
experimental parameters. In detail, in the setup of the experiment in
Paris~\cite{Haroche_RMP,HarocheBook}, the transition frequencies are
$\omega_a = \omega_f = 2\pi\times51\,$GHz, the coupling strength (vacuum
Rabi frequency) amounts to $g=2\pi\times50\,$kHz.  We choose here the interaction times to not exceed $10\,\unit{ms}$, which justifies our assumption of unitary time evolution, being the atomic transition lifetime $30\,\unit{ms}$ and the cavity decay time $0.1\,\unit{s}$. In
particular, we analyze and compare the efficiency of quantum state
preparation for two different time scales: (i) the time scale of the atom-cavity interaction determined by $g$, namely $g\tau\sim 1$, such that $\tau$  ranges in the interval $\tau\sim 10\,\unit{\mu s} - 40\,\unit{\mu s}$ and
(ii) the time scale corresponding to  $g\tau\gg 1$, namely, $\tau=
10\,\unit{ms}$. For such large time scale 
the dressed states of the coupled atom-cavity system are spectrally
resolved up to a precision of the order of $2\pi\times 0.1\,\unit{kHz}$. 

In order to understand the dynamics induced by the optimized controls
and identify the control mechanisms, we  analyze the ground and
excited state populations of the two-level system,
\begin{eqnarray}\label{eq:pop}
\rho_{jj}(t)={\rm Tr}\{\ket{\phi(t)}\bra{\phi(t)}|j\rangle\langle j|\}\,,
\end{eqnarray}
with $j=e,g$, the average photon number in the cavity,
\begin{equation}
  \label{eq:photon}
  \langle n\rangle_{\phi(t)} = \bra{\phi(t)}a^{\dagger}a\ket{\phi(t)}  \,,
\end{equation}
together with its standard deviation,
\begin{equation}
  \label{eq:photonvar}
  \Delta n = \sqrt{\langle n^2\rangle_{\phi(t)}-\langle n\rangle^2_{\phi(t)}} \,,
\end{equation}
and the spectra of the optimized fields $I_\xi(\omega)=| \xi_F(\omega) |^2$, with
\begin{equation}
  \label{eq:spec}
 \xi_F(\omega) = \frac{1}{2\pi}\int_0^{\tau} \tilde\xi(t)e^{-i\omega t}dt \,,
\end{equation}
and  $\tilde{\xi}=\tilde\Omega(t),\tilde\eta(t)$. 

\subsection{Fock states}
\label{subsec:Fock}
We first consider the preparation of the oscillator in a Fock state,
Eq.~\eqref{Fock}, assuming that at $t=0$ it is in the
ground state.   
The preparation of a number state with $n=1$  follows a very simple
dynamics, solely determined by the Jaynes-Cummings Hamiltonian $H_0$ in Eq.
\eqref{Hdrift}~\cite{LawEberly}. In 
fact, by  preparing the atom in the excited state and the cavity
in the vacuum, the cavity will end up in a single photon
$n=1$ Fock state after the  interaction time $\tau=T_0/2$, with
$T_0=2\pi/g$. In general, provided that the initial state is 
$\ket{e,n}$, the Jaynes-Cummings dynamics will naturally let the
system evolve to the state $\ket{g,n+1}$ after the time
$\tau=(2m+1)T_n/2$, with $m$ a natural number and  
\begin{equation}\label{eq:Tn}
  T_n=2\pi/(g\sqrt{n+1})\,.
\end{equation}
This concept requires control of the initial state and of the
interaction time, but no additional field. It is at the
basis of several
protocols~\cite{Domokos,RanganMonroe,Ben-Kish,KneerLaw} for creating 
a Fock state starting from the vacuum
which is relatively simple to prepare. In detail, the
preparation of Fock states with $n>1$ starting from $n=0$ can make use
of external fields which perform $\pi$-pulses on the atom after one
excitation has been transferred to the resonator. The number of such
pulses shall correspond to $n$, such that the interaction time
$\tau\ge \tau_n$ with $\tau_n=\sum_{j=0}^nT_j/2$.    

On the basis of this knowledge, we choose the initial state of the
atom to be $\ket{e}$ and analyze the pulses predicted by OCT for the target
state $\ket{n=4}$ and  an interaction time $\tau=40\,\unit{\mu s}$. 
Note that for the value $g=2\pi\times50\,$kHz, the interaction time $\tau$ we
choose is of the same order of magnitude as $\tau_n$. We
assume that only the atom is driven by an external field, i.e.,
$\tilde\eta(t)=0$, $\Delta(t)=0$, 
since this is relatively simpler to implement in the setup of
Ref.~\cite{Haroche_RMP}. Moreover, amplitude control turns out to be
sufficient, i.e., we keep $\phi_l(t)=0$, and
$\tilde\Omega(t)=\Omega(t)$ is real.  

\begin{figure}[tbp]
  \begin{center}
    \includegraphics[width=0.5\textwidth]{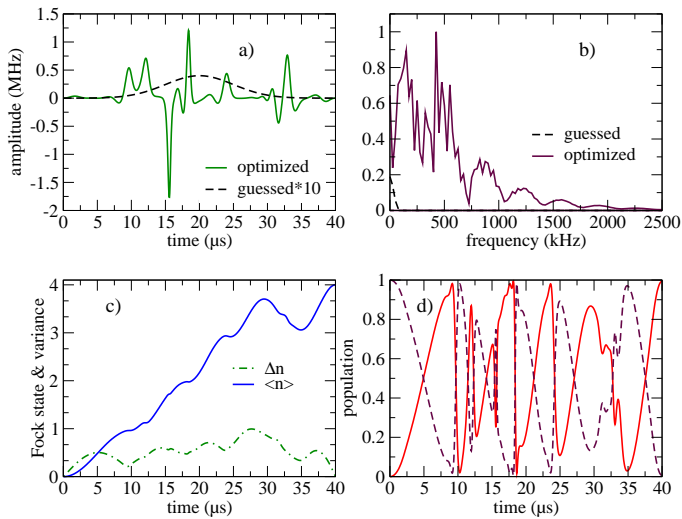}
 \end{center}
  \caption{\label{4_20mus} Preparation of the Fock state $\ket{n=4}$ in
    $\tau=40\,\unit{\mu s}$ using OCT: (a) Amplitude of the pulse driving the atom, $\tilde{\Omega}(t)$, as a function of time (in $\mu$). The guess pulse (dashed line) is a Gaussian with a maximal amplitude of $E_0=40\,\unit{kHz}$ and $\sigma=5\,\unit{\mu s}$.
    (b) Spectrum of the pulse, according to Eq.~\eqref{eq:spec} (since
    $\tilde{\Omega}(t)=\Omega(t)$ is real, the spectrum is symmetric around 0, and we
    show only the positive frequencies). The other subplots show the temporal dependence of 
    (c) mean intracavity-photon number
    $\langle n\rangle$ (solid line) and its standard deviation $ \Delta n$ (dashed line),
    cf. Eqs.~\eqref{eq:photon} and \eqref{eq:photonvar}, and (d)
    population of the atomic states $\rho_{gg}$ (solid line) and $\rho_{ee}$ (dashed line),
    Eq.~\eqref{eq:pop}. The initial state is
   $\ket{e,0}$, the other parameters are specified in the
    text.}
\end{figure}

Figure \ref{4_20mus} displays the temporal (a) and spectral
(b) behaviour of the pulse, which leads to a fidelity $\mathcal
F=1-3\times 10^{-5}$ for reaching the target state. 
The driving field
consists of a series of pulses, whose effect on the dynamics of the
composite system can be inferred from Fig. \ref{4_20mus}(c) and
(d). Figure~\ref{4_20mus}(c) displays the time evolution of the mean
intracavity photon number $\langle n\rangle$ and its standard
deviation $\Delta n$,
cf. Eqs.~\eqref{eq:photon} and~\eqref{eq:photonvar}: The number of 
photons increases in a stepwise manner, like in protocols based on
Refs.~\cite{LawEberly,Domokos}, in which one photonic excitation at
a time is transferred to the cavity mode. Correspondingly, the
variance remains limited, and vanishes exactly at
$t=\tau$. Figure~\ref{4_20mus}(d) shows the populations of the atom's 
ground and excited state as a function of time $t$,
cf. Eq.~\eqref{eq:pop}: They oscillate according to the stepwise
increase of the photon number, performing a series of half Rabi
cycles.  Thus, the protocol obtained numerically essentially 
follows the analytical intuition: Similar to the proposal
of Ref.~\cite{LawEberly} there is only one active coupling at a time.   

At this stage several comments are in order. First, by testing 
other initial atomic states than $\ket{e}$, it turned out that 
the choice of $\ket{e}$ leads to better fidelities for preparing Fock states. The second comment
concerns the interaction time. Optimization 
delivers a pulse that maximizes the fidelity for the specified 
interaction time. Times shorter than $\tau=40\,\mu$s lead to
comparable efficiencies, as long as they are longer than $\tau_n$.
Interaction times shorter than $\tau_n$ lead to inefficient dynamics: 
the fidelity is drastically reduced. We have verified that this 
is a fundamental limit by allowing for optimization of the additional
external fields $\tilde\eta(t)$ and $\vartheta(t)$. These optimizations confirmed that $\tau_n$ is the minimal time, 
needed to transfer one excitation to the cavity containing already $n$ excitations. These considerations are quite generic: We have 
tested them for the preparation of other number states,
$n=2$ and $n=4$, yielding the same conclusion.
Our analysis thus indicates that $\tau_n$ is a
good estimate of the quantum speed limit (namely, the
minimum time required to transfer a quantum state into another,
orthogonal state~\cite{GiovannettiPRA03}) for transferring the vacuum
to the Fock number state $\ket{n}$. This result, moreover, attests to the ability of OCT
to identify the quantum speed limit in cases
where an analytical estimate is not possible~\cite{TommasoPRL09}.
 
We have also verified that it is possible to create other number states with
infidelities of the order of $10^{-5}$ using OCT and solely by 
means of a time-dependent drive on the atom with real amplitude.
We note, moreover, that as there are many possible solutions for the shape of the atom 
pulse, the choice of the guess pulse influences the form of the
optimized pulse but not the final infidelity. The preparation of the Fock state $n=1$ is peculiar since in principle it requires no external control, as the Jaynes-Cummings Hamiltonian naturally drives the atom from the excited to the
ground state by emitting a photon into the resonator with a half Rabi
cycle at the interaction time $\tau_1$. Nevertheless, also in this case and for the same interaction time we find time-dependent dynamics based on external pulses, whose maximum fidelity is the same as the one obtained for the Jaynes-Cummings dynamics without external fields. This property is not surprising \cite{FWM}, and reflects a landscape with several possible optimization. We note, in particular, that the solution with no external field can be indeed recovered by OCT, introducing a cost functional that minimizes the integrated pulse energy. Such a cost
functional, however, leads to inefficient  dynamics for the
preparation of Fock states with $n>1$ as well as for the preparation
of other classes of states.    

\begin{figure}[tbp]
\begin{center}
 \includegraphics[width=0.25\textwidth]{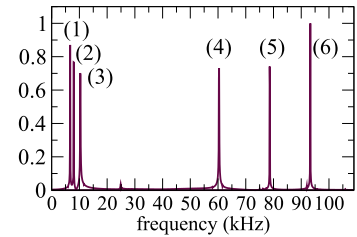}
\end{center}
\caption{\label{4_5ms} Preparation of the Fock state $\ket{n=4}$ in
    $\tau=10\,\unit{ms}$ using OCT: Positive-frequency part of the
    pulse spectrum (in arbitrary units), according to
    Eq.~\eqref{eq:spec} (the spectrum is symmetric about the resonance
    frequency since the amplitude is real). The initial state is
    $\ket{e,0}$, the other parameters are specified in the text. Differing from the other cases, the shape function used for this optimization is a constant. The
    numbers in the plot label transitions between dressed states
    of the atom-cavity system, for details see text.}  
\end{figure}

Even though fast dynamics are in general preferable, the analysis of
the optimization results for longer times is instructive. We now focus on
the preparation of $\ket{n=4}$ starting from $\ket{e,0}$ for the
interaction time $\tau=10\,\unit{ms}$, which is orders of magnitude
larger that $\tau_n$. This interaction time allows for the spectral
resolution of the lowest-energy dressed states of the atom-cavity
system. In this regime, a well-defined number of photons can be pumped
into the resonator by resonantly driving a dressed state of the
Jaynes-Cummings
spectrum~\cite{KimblePhotonBlockade,RempePhotonBlockade}.   
Figure~\ref{4_5ms} displays the spectrum of the optimized pulse which
yields an infidelity of $2\times 10^{-6}$. The amplitude of the
optimized pulse is in general three orders of magnitude smaller than
that in Fig.~\ref{4_20mus} and oscillates on the microsecond time
scale. The spectrum of the optimized pulse in Fig.~\ref{4_5ms}
exhibits well-defined peaks at the dressed states frequencies. In
detail, the peaks which are particularly pronounced correspond to transitions $\ket{\pm,2}\to\ket{\pm,3}$ for (1),  $\ket{\pm,1}\to\ket{\pm,2}$ for (2), $\ket{\pm,0} \to\ket{\pm,1}$ for (3),  $\ket{\pm,0}\to\ket{\mp,1}$ for (4), $\ket{\pm,1}\to\ket{\mp,2}$ for (5), and $\ket{\pm,2}\to\ket{\mp,3}$ for (6). This shows that, in the limit of a sufficiently long interaction time scale, the protocol tends to address the individual transitions between dressed states, adding sequentially excitations till the target state is reached.  

For both $\tau=40\,\mu$s and $\tau=10\,$ms, the required numerical
effort is very moderate: Figure~\ref{4_20mus}
shows the results of an optimization using 4000 iteration steps, with
each iteration taking 5-10$\,$s on a regular workstation computer. For
longer $\tau$ such as in Fig.~\ref{4_5ms} the CPU time per iteration
goes up to 10-20$\,$s but a much smaller number of iterations, of the
order of 100, is required. This indicates that the overall operation
time is a resource for control: The control problem becomes
simpler as more time is available. Note that the exact number of
iterations to reach a certain fidelity threshold depends on the
initial guess. 

\subsection{Fock state superpositions}
\label{subsec:superpos}
\begin{figure}[tbp]
  \begin{center}
    \includegraphics[width=0.49\textwidth]{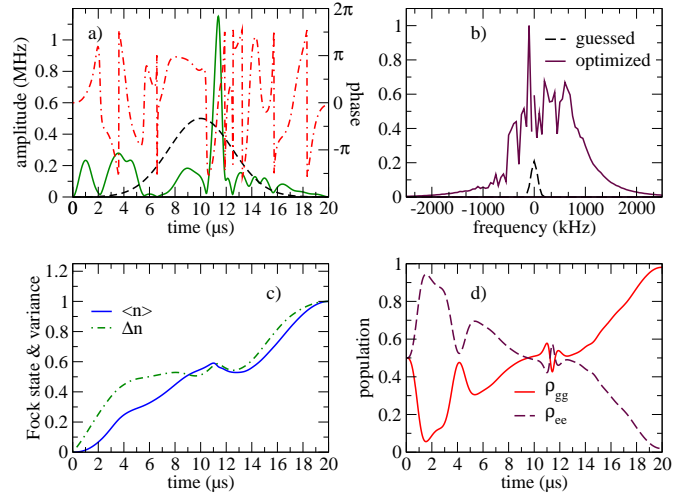}
\end{center}
\caption{\label{0+2_10mus} Preparation of the Fock state superposition
  $(\ket{0}+\ket{2})/\sqrt{2}$ in the interaction time
  $\tau=20\,\unit{\mu s}$ with a pulse driving the atom. (a)
  Temporal dependence of the pulse amplitude (solid line) and phase
  (dashed-dotted line). The dashed line shows the guess pulse
  amplitude (amplified by a factor $10$): This is a Gaussian with a maximal amplitude of $E_0=50\,\unit{kHz}$ and $\sigma=2.5\,\unit{\mu s}$. (b) Spectrum of the pulse, according to Eq.~\eqref{eq:spec}. The amplitude is in
  arbitrary units. The other subplots show the temporal dependence of (c) mean intracavity-photon number $\langle
  n\rangle$ and standard deviation $\Delta n$,and (d) population of
  the atomic states $\rho_{gg}$ and $\rho_{ee}$. The initial state is
  $(\ket{g}+\mathrm{i}\ket{e})/\sqrt{2}\otimes \ket{0}$. } 
\end{figure}

We now analyze the efficiency of preparing a coherent superposition of
Fock states of the form given in Eq.~\eqref{Focksup} using OCT, and
%Following an Law-Eberly type of pulse sequence, Fock state superpositions can be generated from the vacuum state by taking the initial state of the  atom to be a superposition between ground and excited state. The dynamics should then perform a series of half-Rabi cycles, followed by  pulses as the ones employed for the Fock state preparation. 
first discuss in detail the efficiency of preparing the
superposition 
\begin{equation}
  \label{eq:Fock2}
  \ket{\Phi_{\rm target}}=\frac{\ket{0}+\ket{2}}{\sqrt{2}}  
\end{equation}
for the interaction time $\tau=20\,\unit{\mu s}$ and solely using pulses driving the atom. Our study thus focuses on optimizing the field described by the control $\tilde\Omega(t)$ and 
shows that complex-valued $\tilde\Omega(t)$, that is time-dependent amplitude and phase, 
lead to higher efficiencies. Therefore, the
preparation of Fock-state superpositions requires control over both
amplitude and  phase of the pulses. This is a logical consequence of the fact that the final state bears a phase information. Figure~\ref{0+2_10mus}(a) displays the temporal dependence of the pulse, which has been obtained by minimizing the 
functional to a final infidelity of $2\times10^{-5}$. 
The spectrum is shown in Fig.~\ref{0+2_10mus}(b): It is broad and asymmetric about the
resonance frequency. This feature is due to the complex amplitude
of the pulse.  Figure~\ref{0+2_10mus}(c) shows that  the intracavity photon number
evolves from the vacuum state to the desired target state, which is
characterized by a mean photon number $\langle n\rangle=1$ and
standard deviation $ \Delta n=1$. The atomic level populations in
Fig. \ref{0+2_10mus} (d) display  
some oscillatory behaviour until almost all the population ends up in $\ket{g}$. We note that better
efficiencies are found by taking the initial atomic state to be a
superposition of ground and excited state: In the case of
Fig. \ref{0+2_10mus} the initial atomic state is $\ket{\Psi_{
  a}}=(\ket{g}+{\rm i}\ket{e})/\sqrt{2}$. 
      
\begin{figure}[tbp]
  \begin{center}
    \includegraphics[width=0.25\textwidth]{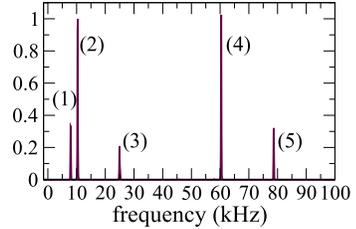}
    \caption{\label{0+2_5ms} Preparation of the Fock state
      superposition $(\ket{0}+\ket{2})/\sqrt{2}$ for the interaction
      time $\tau=10\,\unit{ms}$. Positive-frequency part of the
      spectrum of the pulse (in arbitrary units), according to
      Eq.~\eqref{eq:spec} (since the pulse is real, the spectrum is
      symmetric about the resonance frequency). The initial 
      state is $(\ket{g}+\mathrm{i}\ket{e})\otimes\ket{0}/\sqrt{2}$. The shape function used for this optimization is a constant. The numbers label transitions between dressed states, for details see text.
    }  
  \end{center}
\end{figure}

When instead the interaction time is large, $\tau=10\,\unit{ms}$, the
amplitude of the pulses becomes approximately three orders of magnitude
smaller in comparison to the pulses obtained for shorter interaction
times. Here, the amplitude changes are carried out on time scales of the order of microseconds. 
Figure~\ref{0+2_5ms} displays the spectrum of the
corresponding optimized pulse, which leads to a final infidelity
below $10^{-10}$. Since the pulse is real, its form is symmetric about the resonance frequency. The peaks correspond to the frequencies of transitions between dressed
states of the atom-cavity system, as follows:
(1) corresponds to the transition
$\ket{\pm,1}\to\ket{\pm,2}$, (2) to $\ket{\pm,0}\to\ket{\pm,1}$,
(4) to $\ket{\pm,0}\to\ket{\mp,1}$ and (5) to
$\ket{\pm,1}\to\ket{\mp,2}$, whereas (3) corresponds to the
transition from the cavity vacuum state to the dressed states
$\ket{\mp,0}$.
The spectrum shows that, for sufficiently long
interaction times, quantum-state preparation here is reached by means
a pulse which resonantly drives the dressed states 
transitions leading to the target state.  
	
\begin{figure}[tbp]
  \begin{center}
    \includegraphics[width=0.99\linewidth]{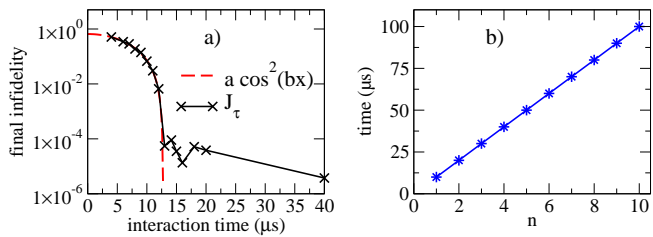}
    \caption{\label{focksup_time} (a) Final infidelity $J_{\tau}$ as a
      function of time (in $\mu s$) for the preparation of the target
      state $(\ket{0}+\ket{2})/\sqrt{2}$. The crosses correspond to
      the numerical data, the solid line serves as a guide for the
      eyes, whereas the red dashed line corresponds to the fitted curve
      $f(x)=a\cos^2(bx)$, with $a=0.25$ and $b=0.66$. (b) Minimum time
      (in $\unit{\mu s}$)  as a function of $n$ required to prepare the target 
      state $(\ket{0}+\ket{n})/\sqrt{2}$ with infidelity $<10^{-4}$. The initial state is  
      $(\ket{g}+\mathrm{i}\ket{e})\otimes\ket{0}/\sqrt{2}$, and only the atom-pump
      pulse $\tilde{\Omega}(t)$ is employed in Eq.~\eqref{Hctrl}.}  
  \end{center}
\end{figure}

We finally perform an optimization over the minimum time required to
prepare the superposition  
\[
\ket{\Phi_{\rm target}}_n=\frac{\ket{0}+\ket{n}}{\sqrt{2}}
\]
as a function of the Fock-state number $n$. In doing so, we require
the infidelity to be below $10^{-4}$. We take the initial state to be 
\[
\ket{\phi(t=0)}=\frac{\ket{g}+\mathrm{i}\ket{e}}{\sqrt{2}}\otimes \ket{0}\,,
\]
and optimize only the atom pump pulse
$\tilde\Omega(t)$. Figure~\ref{focksup_time}(a) displays the final
infidelity $J_{\tau}$ as a function of the interaction time $\tau$ for
the preparation of the state $(\ket{0}+\ket{2})/\sqrt{2}$. The
behaviour clearly shows that  infidelities below $10^{-4}$  are
reached for interaction times exceeding $12\,\unit{\mu s}$. At shorter
times, the infidelity decreases with a functional behaviour that can
be fitted to the function $f(x)=a\cos^2(bx)$. Analogous behaviour has
been reported in other optimization
studies~\cite{TommasoPRL09,Egger}. The sharp change in the
infidelity 
allows for the determination of the quantum speed limit, which for
this specific state and dynamics lies at about $12\,\unit{\mu s}$. 
By means of the same  procedure we identify the minimum time required
to prepare the state $(\ket{0}+\ket{n})/\sqrt{2}$ 
with final infidelity below $10^{-4}$ as a function of $n$. 
Figure~\ref{focksup_time}(b) shows that the required
interaction time scales linearly with $n$. This time exceeds the time
found by simple estimates, which should scale as $\sqrt{n}$ for
protocols which make use of a third, ancillary state as in Ref. \cite{Wilhelm}. Here, instead,
the scaling of time is found for the evolution constrained to occur
between the two states $|g\rangle$ and $|e\rangle$ coupling resonantly
to the cavity field.  

We note that the preparation of Fock state superpositions is achieved
here by solely employing an external pulse that drives the atom. We
have also applied optimization of the pulse driving the cavity and
that generating 
a time-dependent dynamical Stark shift on
the atom, but found no significant improvement with respect to the
case in which only the pulse on the atom was employed. A systematic
comparison indeed reveals that the atom pulse leads to the target
state in an efficient way, whereas employing solely the pump on the cavity leads to
dynamics which are characterized by lower fidelities over time scales
of the order of tens of microseconds. The required numerical effort
is very similar to the one reported in Section~\ref{subsec:Fock}.

\subsection{Schr\"odinger cat states}
\label{subsec:cat}

In this section we focus on the deterministic preparation of
Schr\"odinger cat states, cf. Eq.~\eqref{cat}. They are also known in the
literature as ``even cat states'' since their decomposition in the
Fock state basis only contains even number states. We target $\alpha=
1+\mathrm{i}$, which is sufficient for a proof-of-principle discussion.
Realizing larger values of $|\alpha|$, in fact, requires a significantly larger numerical
effort since the Hilbert space of the cavity has to be truncated at much
larger Fock state number $n$. Unlike for the previous
classes of states, efficient preparation for the considered time scale requires a 
pulse that directly drives the resonator. This condition can
intuitively be understood since a cavity pump efficiently generates
coherent states of the oscillator.  For the results reported in this
section, the dynamics employs both a pulse driving the resonator, 
$\tilde{\eta}(t)$, and a pulse which couples to the atomic
transition, $\tilde{\Omega}(t)$.  

\begin{figure}[tbp]
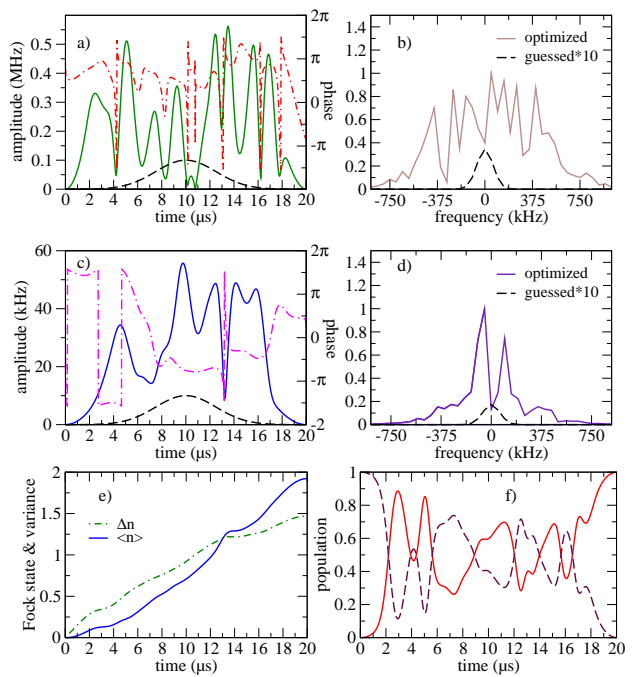

  \begin{center}
    \includegraphics[width=0.93\linewidth]{Fig7a}
    \hspace*{2.5ex}\includegraphics[width=0.93\linewidth]{Fig7b}
    \caption{\label{cat_10mus} Preparation of the even cat state with
      $\alpha = 1+\mathrm{i}$ in $\tau=20\,\unit{\mu s}$ using OCT: Temporal
      dependence of the optimized pulses driving the atom (a) and the
      cavity (c). The solid and dashed-dotted line correspond to the
      amplitude and the phase of the fields, respectively (the dashed line shows
      the guess-pulse amplitude amplified by a factor $10$, which is a Gaussian pulse with $E_0=10\,\unit{kHz}$ and $\sigma=2.5\,\unit{\mu s}$ for the atom and with $E_0=1\,\unit{kHz}$ and $\sigma=2.5\,\unit{\mu s}$ for the cavity.  Subplots (b)
      and (d) display the corresponding spectra. The other subplots show the temporal dependence of
      (e) mean intracavity-photon number $\langle
      n\rangle$ and standard deviation $ \Delta n$ and (f) population of
      the atomic states $\rho_{gg}$ (solid line) and $\rho_{ee}$ (dashed
      line). The initial state is $\ket{e,0}$, the
      other parameters are given in the text. }  
\end{center}
\end{figure}
Figure~\ref{cat_10mus} displays the resulting pulses obtained using
OCT taking the initial state $\ket{e,0}$ and the interaction time
$\tau=20\,\mu$s. The final infidelity here amounts to
$6\times10^{-4}$. We observe that the preparation of the even cat
state with $\alpha = 1+\mathrm{i}$ requires complex pulses, both for
atom (Fig.~\ref{cat_10mus}(a)) and cavity
(Fig.~\ref{cat_10mus}(b)). Moreover, the amplitude of the atom pump
pulse is about one order of  magnitude larger than the vacuum Rabi
splitting (which scales with $g$) and than the amplitude of the cavity
pump pulse. This suggests that the field $\tilde{\Omega}(t)$ dresses
the atomic levels, whereas the cavity pulse drives selectively the
resonances of the dressed atom. This intuition is corroborated by the
spectrum of the pulse driving the cavity, shown in
Fig.~\ref{cat_10mus}(d), where  
two prominent peaks appear at the transition frequencies between the dressed states. 
Further information is extracted from the mean photon number in
Fig.~\ref{cat_10mus}(e) and 
corresponding variance. They show a steady increase, as it would correspond
to a displacement of a harmonic oscillator whose amplitude increases
with time until the target value is reached. The atomic dynamics,
Fig.~\ref{cat_10mus}(f),  
exhibits oscillations on the time scale of a few microseconds. This is
consistent with the picture of the atom pulse dressing the atomic
transition. These features thus suggest that the system undergoes a
conditional dynamics, such that the cavity pulse drives the two resonances
of the dressed atom. This dynamics is reminiscent of proposals for the
preparation of Schr\"odinger cat states based on projective measurements~\cite{SolanoAgarwal}, 
in which even or odd cat states are prepared by means of a conditional dynamics, which realizes 
a displacement of the oscillator whose sign depends on the internal state of the dressed atom. Differing from these
proposals, the preparation presented here is deterministic 
and thus does not rely on a final projective measurement.

\begin{figure}[tbp]
  \begin{center}
    \includegraphics[width=0.49\textwidth]{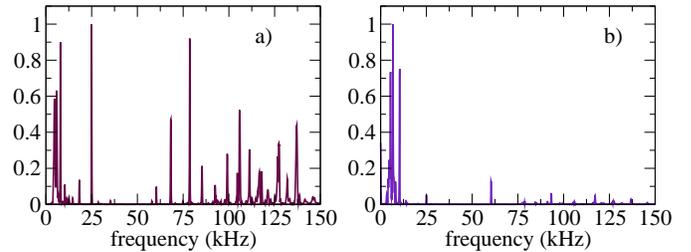}
    \caption{\label{cat_5ms} Preparation of the even cat state with
      $\alpha = 1+\mathrm{i}$ in $\tau=10\,\unit{ms}$ using OCT:
      Positive-frequency spectra of the pulse driving the 
      atom (a) and the cavity (b). The amplitude is in arbitrary
      units. The initial state is $\ket{e,0}$, the other parameters are
      given in the text.}  
  \end{center}
\end{figure}

We now analyze the efficiency of preparing an even
cat state with $\alpha = 1+\mathrm{i}$ for an interaction time of 
$\tau=10\,\unit{ms}$ using OCT. The atom and cavity pulses turn out to be
complex. Differing from the previous case, however,  their amplitudes
are now comparable, and three orders of magnitude smaller than the
atom pulse in Fig.~\ref{cat_10mus}(a). 
Figure~\ref{cat_5ms} displays the spectra of atom and
cavity pulse for a final infidelity of $8\times10^{-5}$. The spectra
show well defined resonances at the dressed states of the
Jaynes-Cumming dynamics, where the number of states addressed by the
fields is now significantly larger. This is consistent with the fact 
that the ideal cat state corresponds to an infinite sum over the Fock
number states. A striking feature is  
the relative weight between the spectral lines of the atom and the
cavity pulse: The latter has a few predominant contributions at low
frequencies, corresponding to the lowest dressed states of the
ladder. These features seem to suggest that the cavity pump drives
selectively and coherently the individual dressed state such that the
resulting superposition delivers the target state~\eqref{cat}. 

Interestingly, the number of OCT iterations required to reach a
certain fidelity is independent of the operation time for cat
states. This is strikingly different from the preparation of Fock
states and Fock state superpositions where time appears to be a
resource for control. We attribute this difference to the fact that an
infinite sum over Fock states is required for an ideal cat state which
cannot perfectly be achieved in either one of the interaction times
employed in our optimizations.

\section{Accounting for experimental uncertainties and noise}
\label{sec:exp}

The discussion so far has been concerned with the preparation of
high-fidelity quantum states for precisely known interaction times:
The pulses which we have determined lead to efficient dynamics as long
as the initial and final time of the interaction are well
defined. Moreover, we have assumed  the 
parameters of Hamiltonian~\eqref{Hrichtig} to be precisely known. 

In this section we investigate the efficiency of the state preparation
when taking parameter fluctuations and other sources of
uncertainty into account during the optimization. For this purpose we
consider the specific situation of the experiment at ENS \cite{Haroche_RMP,HarocheBook}. We
systematically include in our optimization procedure the following
experimental features that might compromise our control protocols: (i) the
position-dependent coupling of the atom with the cavity mode. In
particular, the parameter $g$ in Hamiltonian~\eqref{Hdrift} is
replaced by the position-dependent function $g\to g(x,y,z)$, where $x$
is the propagation direction of the atom with 
constant velocity $v$, $x=v t$, and $y,z$ are the transverse axes. The
functional form reads 
\begin{equation}
  g(x,y,z)= g_0 {\rm e}^{-\frac{(x-x_0)^2+(y-y_0)^2}{\sigma^2}}\cos(2\pi(z-z_0)/\lambda)\,,
\end{equation}
where $x_0$ and $y_0$ refer to the cavity center in the transverse
plane, $z_0$ to the closest antinode, $\lambda=5.87\,\unit{mm}$ is the
cavity-mode wavelength, $\sigma=6\,\unit{mm}$ is the mode waist at the
cavity center, and $g_0=2\pi\times50\,\unit{kHz}$ . We further account for the
fact that (ii) the atom can 
only be localized up to a precision of $1\,\unit{mm}$ in each direction.
This gives rise to a fluctuation in the value of $g$ as well as to
fluctuations of  the initial and final times of the interaction. In
addition, due to the  geometric properties of the present cavity-QED
setup, (iii) a transverse pulse, $\Omega(t)$, can also directly couple to the
cavity mode. Roughly, a $\pi$-pulse on the atom
pumps about 20 photons into the cavity mode.  Moreover, 
(iv) temporal fluctuations of the cavity mode frequency occur which
can vary up to  $\pm 5\,\unit{kHz}$ in a day. Finally, (v) technical 
problems in the generation of the pulses, such as digitalization and
finite time response of the pulse generator, need to be considered. 

In our analysis of the optimized solutions' performance we choose the interaction time to be
$\tau=\sigma/v \sim 100\unit{\mu s}$  when taking
these effects into account. This corresponds to a minimal
velocity of the atoms of about $v\sim 60\unit{m}/\unit{s}$. We set the threshold for efficient state preparation by requiring
infidelities of the order of $10^{-2}$. This corresponds to the
present state-of-the-art of the experimental quantum-state 
discrimination~\cite{Sayrin}.

Our optimization strategy accounts for these effects by optimizing an
ensemble of trajectories whose dynamics are governed by Hamiltonians
with different system parameters, but which experience the same
control fields~\cite{Goerz,Kobzar2004,Kobzar2008}. Effects (i) and (ii) are implemented
by varying the position-dependent coupling $g(x,y,z)$. Specifically, 
$g(x,y,z)$ is evaluated for three values of $x$, $y$ and $z$,  namely,
$x=x_0,x_0\pm0.5 \unit{mm}$, and analogously for $y$ and $z$. Effect
(iii) is simulated by the additional Hamiltonian term
\begin{equation}
  \label{Hundesired} 
  H''(t) = \hbar \xi \frac{\Omega(t)}{2}(e^{-i(\phi_l(t)+\omega_l
    t)}a^{\dagger}+{\rm H.c.})\,, 
\end{equation}
where $\xi$ is a geometric coefficient deduced from the experiment, 
which can take a value between 1 and 4. We consider the worst case and
fix the coefficient to take  mean value $\xi=4$, with a variation
of $\pm$10\%. Such a precision in determining $\xi$
could be accessible in an accurate auxiliary measurement using a direct counting of the photon number in the cavity \cite{Guerlin}. The variation in the cavity frequency, effect (iv), is
accounted for by varying the corresponding parameter in
Eq. \eqref{Hrichtig}: 
specifically, we consider the three values $\omega_f
=\omega_{f,0},\omega_{f,0}\pm 2\pi\times 5\unit{kHz}$. Optimization then consists
in propagating the same initial state with different Hamiltonians, in
order to identify the control fields which yield the best average
fidelity. A different approach is required for simulating the effect
of digitalization and finite-time response, effect (v). The effect of digitalization is
accounted for by fixing the time-step size for changes in the field to
$100\,\unit{ns}$. Additionally, we model the imprecision and the
response of the pulse generator by adding white noise to the atom pulse, with a maximal
amplitude of $2\pi\times 1\,\unit{kHz}$. Two realizations of
the random noise are considered during the optimization. 
The resulting ensemble
comprises 324 system copies when all effects are taken into
account. The average fidelity 
is determined by integrating over the complete parameter ranges. 

We first check how much these effects deteriorate the fidelities when
not taken into account during the optimization.  
We consider the target state in Eq.~\eqref{eq:Fock2}, and determine
the infidelity for the initial state
$(\ket{g}-\mathrm{i}\ket{e})\otimes\ket{0}/\sqrt{2}$. 
While the infidelity for the preparation of the target
state under ideal conditions is $1.0\times10^{-4}$, with the same
optimized pulse the infidelity is increased to $9.97\times10^{-1}$ when
the experimental imperfections are accounted for. It is thus
imperative to include the imperfections in the
optimization if one is to provide pulses that will be meaningful in
the experiment. 

\begin{table}[tb]
  \centering
  \begin{tabular}{|l|c|c|c|}\hline
   noise effect & average &size of & average \\[-0.5ex]
                           & optimization  &optimization & integrated \\[-0.5ex]
                           & infidelity & ensemble& infidelity \\
    \hline
    %uncertainty in coupling $g$, 
    (i)-(ii)  & $5.0\times 10^{-3}$&27 & $4.6\times 10^{-3}$\\ 
    %cavity excitation due to atom-pump pulse, 
    (iii) & $2.2\times    10^{-2}$&2 & $1.3\times 10^{-2}$\\ 
    %(iii) & $4.8\times     10^{-2}$&10 & $4.1\times10^{-2}$\\ 
   %uncertainty of cavity mode frequency, 
    (iv)  & $2.1\times 10^{-4}$&3 & $1.6\times 10^{-4}$\\
   %amplitude noise, 
    (v) &  $6.2\times 10^{-5}$&2 & $1.2\times 10^{-4}$\\ \hline
  \end{tabular}
  \caption{Classification of noise impact: (i)-(ii) uncertainty in
    atom-cavity coupling $g$, (iii) undesired cavity excitation due to atom pulse,
    (iv) uncertainty of cavity mode frequency, (v) amplitude noise due
    to digitalization and finite time response of pulse
    generator. Average optimization fidelity refers to the final value
    of $J_\tau$ evaluated for the system copies in the optimization
    ensemble, whereas 
    integrated average fidelity corresponds to an integral over the
    parameter range. The single contributions are to be compared to
    an average optimization infidelity of $1.7\times 10^{-1}$ (for an
    ensemble of 324 system copies) and an
    average integrated fidelity of $1.6\times 10^{-1}$  when all noise
    sources are accounted for simultaneously.
  } 
  \label{tab:class}
\end{table}

% \begin{table}[tb]
%   \centering
%   \begin{tabular}{|l|c|c|}\hline
%    noise effect & minimal &ensemble  \\
%                            & infidelity after&size \\
% 		& propagation&\\
%     \hline
%     %uncertainty in coupling $g$, 
%     (i)-(ii)  & $1.9\times 10^{-3}$&125\\ 
%     %cavity excitation due to atom-pump pulse, 
%     (iii) & $2.1\times
%     10^{-1}$&100\\ 
%    %uncertainty of cavity mode frequency, 
%     (iv)  & $2.4\times 10^{-4}$&100\\
%    %amplitude noise, 
%     (v) &  $6.5\times 10^{-5}$&100\\ \hline
%   \end{tabular}
%   \caption{Classification of noise impact.}
%   \label{tab:class2}
% \end{table}
When all experimental uncertainties and  noise sources are accounted
for, using ensemble OCT, 
the optimized pulses show an improvement by an order of magnitude,  
leading to an infidelity of about $1.6\times 10^{-1}$. More
importantly, optimization in the presence of noise and imperfections
also allows us to identify 
whether this infidelity is caused by several factors simultaneously or
whether it is mainly due to a single source. This is analyzed in
Table~\ref{tab:class} which reports 
the individual contributions of each noise effect to the total
infidelity of $1.6\times 10^{-1}$.
The optimizations were taken to be converged
when $J_\tau$ changed by $10^{-6}$ or less.
%%% chr: der folgende Satz kann auch auskommentiert werden, was ich
%%% damit andeuten moechte, ist, dass man vermutlich noch weiter
%%% kommt, wenn man herausfindet, welche 2 Effekte sich negativ
%%% verstaerken (meine Vermutung: ii und iii). Dann koennte man mit
%%% ii-iii simultan voroptimieren, bevor man alle 324 Kopien loslaufen
%%% laesst. bin mir aber nicht sicher, dass wir das jetzt noch tun
%%% wollen... 
The fact that the largest infidelity of a single noise effect in
Table~\ref{tab:class} is about one order of magnitude smaller than the
infidelity when accounting for all effects simultaneously, 
$1.3\times 10^{-2}$ for effect (iii) compared to $1.6\times 10^{-1}$,
suggests that there is an interplay between two or more of the noise
effects that increases the difficulty of the optimization problem
substantially. 
The  largest impact on the fidelity, as clearly revealed by inspection
of Table~\ref{tab:class},  is due to
undesired excitation of the cavity by the atom pump pulse. 
%%% chr: den folgenden Text verstehe ich nicht, was ist mit 'driving
%%% from the side' gemeint? 
%%% Warum wuerde man erwarten, dass Optimierung des Atom-Pulses den
%%% Effekt des Cavity-Pulses aufhebt? Das haengt ganz sicher von der
%%% Optimierungszeit ab (zusaetzlich zur Abhaengigkeit von g)
Indeed, we have tested that efficient preparation of  Fock state
superpositions is achieved by pulses which drive the atom from the
side, while large pulses on the cavity field tend to decrease the
efficiency over short times. Additionally, in this specific situation, 
the cavity pulse is proportional to the pulse of the atom, and its
effect thus cannot  simply be suppressed by better optimization of the
atom pulse. We have verified that the fidelity is partially improved by
taking a larger value of the vacuum Rabi frequency,
$g_0=2\pi\times100\,$kHz, even for shorter time scales and
$\tau=50\,\mu$s. More could be achieved by applying a pulse on the
cavity field whose role is to suppress the detrimental effect of
Eq.~\eqref{Hundesired}. A simpler possibility would be to reduce the parasitic coupling, for instance by proper engineering of the microwave field map.

In the current work, we have not considered systematic errors in the control parameters $\Omega(t)$ and $\eta(t)$. They may be due to our imperfect knowledge of the real field amplitudes coupled to the atom and the cavity. In contrast to noise, systematic errors may have cumulative effects and lead to faster deterioration of the fidelity of state preparation. Taking them into account in OCT might give preference to solutions featuring adiabatic passage elements instead of resonant population transfers. In this case it is expected that efficient optimization shall resort to the control parameter $\Delta(t)$. We plan to continue the study of this subject in more detail.

\section{Conclusions}
\label{Sec:conclude}

We have employed optimal control theory to identify dynamics
that lead to efficient quantum state preparation of a harmonic
oscillator, using a single atom as ancillary quantum system. The
pulses which were optimized 
pump either the atomic transition or the cavity field, or both. 
The temporal shape of their amplitudes and phases
have been determined by optimization using Krotov's
method.  We have focused on the efficient preparation of three classes of
quantum states which have previously been discussed in the
literature. This choice has enabled us to 
compare our results to the dynamics constructed from a physical
understanding based on the quantum optical master
equations. Our optimized pulses for the preparation of 
Fock states, Fock state superpositions and Schr\"odinger cat states
yield errors below $10^{-4}$, provided that all parameters of the
Hamiltonian are precisely known. Optimal control theory has also
allowed us to determine the minimum interaction time required for
the quantum state preparation. As expected, it is determined by
the atom-cavity interaction strength and the type of state, scaling
for example linearly with $n$ for a superposition of the type
$(\ket{0}+\ket{n})/\sqrt{2}$.

We have furthermore evaluated the efficiency of optimal-control-theory-based
protocols for prospective applications by taking into account
the specific experimental conditions of the setup at ENS. Our analysis
shows that it is crucial to include parameter uncertainties and
fluctuations in the model. When the noise sources are accounted for during the
optimization, optimal control theory can counteract
their detrimental influence. This typically improves the error by at
least  one order of magnitude. Even more importantly, optimal control theory also
allows for identifying the noise source with the largest detrimental
impact.
% For the example of the ENS setup, this turned out to be the unknown coefficient in the undesired interaction of the cavity mode with the atom pump. If this coefficient were more precisely known, the error could easily be pushed below the current state discrimination capability without the necessity of improving the experimental setup. 
When experimental imperfections are properly taken into account, 
optimal control theory can provide an efficient route to
quantum-state preparation of \textit{arbitrary}
states. The identified protocols solely rely on deterministic dynamics
and are indeed efficient over realistic time scales and for
state-of-the-art experimental conditions.  

Our protocols can be straightforwardly applied to
other physical systems with similar features, for instance, 
for quantum-state preparation in circuit QED, of the quantized motion of a trapped
atom or ion, and of the quantized vibrations of a micromechanical
resonator. The robustness against parameter fluctuations has to be
calibrated to the specific experiments, but the perspectives are 
similarly promising to the ones found for microwave cavity QED. 

We finally observe that the dynamics implemented here require an
efficient determination of the initial state, which has to be uniquely
defined. One could also relax this condition and consider
implementations which merge optimal control theory with concepts like quantum reservoir
engineering~\cite{QRE,QRE2,QRE3,Christian} to realize robust
quantum-state preparation~\cite{DanielNJP13}.
    
\begin{acknowledgments}
We are grateful to Michel Brune, Daniel
Egger, Stefan Gerlich, and in particular to Dietrich Leibfried and Frank Wilhelm, for discussions and helpful comments. Financial support from the German Federal Ministry of Education and Research (BMBF, project
QuOReP 16BQ1011), by the European Commission (IP AQUTE),  and by the
German Research Foundation (DFG) is acknowledged. K. R. acknowledges
supports from the StudienStiftungSaar and the kind hospitality of the
microwave cavity QED group at ENS in Paris. G.M. thanks the Ion Storage
Group at NIST (Boulder) for hospitality during completion of this work.
\end{acknowledgments}

\bibliography{cavity_oct}

\end{document}